\documentclass[aps,pre,twocolumn,showpacs,a4paper]{revtex4}
\usepackage{amssymb,amsmath}
\usepackage{epsfig}
\usepackage{graphicx}
\usepackage{url}
\usepackage{subfigure}

\begin{document}

\title{Density-feedback control in traffic and transport far from equilibrium}

\author{Marko Woelki\footnote[1]{Marko.Woelki@DLR.de}}

\affiliation {Institute of Transportation Systems, German Aerospace Center, Rutherfordstra{\ss}e 2, 12489 Berlin, Germany}

\begin{abstract}
A bottleneck situation in one-lane traffic-flow is typically modelled with a constant demand of entering cars. However, in practice this demand may depend on the density of cars in the bottleneck. The present paper studies a simple bimodal realization of this mechanism to which we refer to as density-feedback control (DFC): If the actual density in the bottleneck is above a certain threshold, the reservoir density of possibly entering cars is reduced to a different constant value. By numerical solution of the discretized viscid Burgers equation a rich stationary phase diagram is found. In order to maximize the flow, which is the goal of typical traffic-management strategies, we find the optimal choice of the threshold. Analytical results are verified by computer simulations of the microscopic TASEP with DFC.
\keywords{TASEP, Burgers equation, traffic management, toll}
\end{abstract}
\pacs{05.40.-a, 05.70.Ln, 02.50.-r}

\maketitle

\section{Introduction}
In the physical literature, traffic flow is modelled from different viewpoints as hydrodynamic models (on a macroscopic scale) or microscopic stochastic models. Microscopic approaches usually can be considered as a generalization of the so-called totally asymmetric exclusion process (TASEP). The model is defined on a discrete one-dimensional lattice that represents the road. Each lattice site can either be empty or occupied by exactly one particle (car). If the site in front is empty, cars move to the next site at a certain rate or probability depending on the dynamics (either random-sequential or parallel). This process is widely studied mathematically and due to its exact solvability it is of great interest for non-equilibrium statistical physics, see \cite{Mallick_review} for a recent review. Of particular mathematical interest is the model with open boundaries, where particles may enter the first site at rate $\alpha$ and leave the last site at rate $\beta$ that differ from the bulk-hopping rate in general. Depending on the values of those parameters one finds that the system can be in either of three phases, a low-density phase, a high-density phase and a maximum-current phase. For traffic applications one often uses a parallel update instead of the generic random-sequential update studied here. Note that if cars are allowed to move further than a single site under such a parallel update scheme this leads to the so-called Nagel-Schreckenberg model \cite{nasch}. On the other hand, the macroscopic approaches are typically based on investigations of Lighthill and Witham \cite{light} who described the effect of moving traffic jams by travelling-wave solutions of a simple partial differential equation. Since this inital work, there is a number of generalizations of the hydrodynamic approach \cite{nagel}, \cite{helbing}. For example the viscid Burgers equation is a generalization of the Lighthill-Witham equation with an additional diffusive term. This modification is enough to describe qualitatively on hydrodynamic Eulerian scale the TASEP phase diagram, see \cite{Blythe} for further references. By discretization of space, Burgers equation recovers the mean-field equations of the TASEP in which correlations between neighboring sites of the lattice are neglected \cite{Pop}.

The present paper models a road section to which cars can enter at the left end and leave at the right end. Common physical approaches of microscopic and macroscopic models assume a constant demand for entering the lattice. In the TASEP this is reflected by a constant rate $\alpha$ at which a particle enters the first lattice site if it is empty. From the viewpoint of Burgers equation this corresponds to a constant left reservoir density $\rho_l = \alpha$ of customers. This fact will be changed in our investigations, see \cite{Wood}, \cite{Adams}, \cite{Cook08}, \cite{Cook} for related approaches. One way to think about it is to assume that those customers have a route alternative \cite{Cook12}, \cite{Fosgerau}, \cite{brankov} and that they can anticipate the density of cars on the road section, then a fraction of those customers will take an alternative if the density $\rho$ exceeds a certain threshold $\rho^*$. Thus the density of potential customers is reduced from $\rho_l = \rho_-$ to $\rho_l = \rho_+$ if $\rho > \rho^*$. In TASEP, this change of the reservoir density is reflected by different insertion rates $\alpha_- = \rho_-$ and $\alpha_+ = \rho_+$. The same scenario can be transferred from the viewpoint of individual drivers to the viewpoint of a traffic-management center that tries to control the density in the system in order, for example, to maximize the flow. At both ends of the road section there might be sensors that count entering and leaving cars and the controller is able to change the inflow if a certain number of cars is exceeded.
Obviously if one does not control the outflow from the bottleneck as well, one will not generally be able to keep a desired density in the system. However, it is interesting to decide whether this incomplete regulation can be appropriate for real traffic situations in certain parameter regions. The scenario can be interpreted as a sort of ramp-metering and reflects a common way of flow maximization in practice \cite{kerner}, \cite{helbing_ramp}, \cite{Markos2}. One way to reduce the time-averaged inflow is by a traffic light that switches the effective left-reservoir density to zero from time to time \cite{Pop,Neumann,Barlovic}. Another possible application of this varying input rate is the concept of dynamic toll: At the entrance (which plays the roll of a toll booth) a prize for passing the road section is computed in dependence of the current occupation of vehicles \cite{RampHOV}, \cite{Markos}, \cite{Fosgerau}.
While those problems are specially dedicated to traffic, the considerations of the present paper are quite general so that results apply not only to traffic but to other transport scenarios far from equilibrium (see \cite{Blythe} for an overview of applications in other research areas as intra-cellular transport) with density-feedback control as well.\\
The remainder of the paper is as follows. In section \ref{s1}, we define the TASEP with density-feedback control (DFC) that generalizes the particle-insertion procedure of the usual TASEP. We continue by deriving its mean-field equations from Burgers equation with modified boundary condition. The following section \ref{s2} presents analytical results from numerical solutions of the mean-field equations. Special interest is given to the phase diagram of the TASEP influenced by DFC. Section \ref{s3} shows how DFC can be used for flow optimization in TASEP and highlights the benefit of DFC in contrast to the generic TASEP. In section \ref{s4} computer simulations of TASEP with DFC are presented and compared to the analytical predictions, before we formulate our conclusions.
\section{Model definitions}
\label{s1}
First, the mechanism of density-feedback control is defined from the microscopic and macroscopic viewpoint and shown how they translate into each other.
\subsection{Density-feedback control TASEP}
The microscopic TASEP model is defined on a one-dimensional lattice with $L$ sites, labeled from left to right as $l = 1$, $2$, $\dots$, $L$. Each site is either occupied by a single particle or is empty; this defines its time-dependent states $\tau_l(t) = 1$ (occupied) and $\tau_l(t) = 0$ (empty). Particles whose right neighboring site is vacant may move onto this site at rate $p$. From the last site a particle leaves the system at constant rate $\beta$, while particles enter the system on site 1 at rate $\alpha$. The process is considered in continuous time, where we can set the time scale by taking $p=1$.
In the following we consider the TASEP with density-feedback control (DFC) which implies modified particle insertion:
\begin{eqnarray}
\alpha(N)= \begin{cases} \alpha_-, \text{ for } N < N^*\\ \alpha_+, \text{ for } N \geq N^*.\end{cases}
\end{eqnarray}
Hence the probability that a particle enters the lattice at site 1 takes a different value if the actual particle number $N$ is above or below a threshold $N^*$.\\
We note certain limits of this process: if $\rho^*=0$ ($\rho^*=1$) one recovers the TASEP with $\alpha = \alpha_+$ ($\alpha = \alpha_-$). If we take $\alpha_+=0$ the process is very related to the works of references  \cite{Adams}, \cite{Cook08}, \cite{Cook}. In those works, however, the TASEP is considered with a constrain on the overall particle number including the single reservoir from which particles are injected and to which particles leave the lattice.
\subsection{Burgers-equation approach}
The starting point for the macroscopic description is the viscous Burgers equation
\begin{equation}
\frac{\partial \rho}{\partial t} + \frac{\partial(\rho(1-\rho))}{\partial{x}} = D\frac{\partial^2 \rho}{\partial x^2},
\end{equation}
for the density $\rho = \rho(x,t)$ with the right boundary condition $x(L,t) = \rho_r$. Instead of the generic left-hand boundary condition
\begin{equation}
\label{generic_bc}
\rho(0,t) = \rho_l
\end{equation}
we use a dynamical density $\rho_l(t)$ that depends on the (spatially) averaged density $\bar{\rho}(t)$ at time t as
\begin{eqnarray}
\label{dynamic_bc}
\rho_l(t)= \begin{cases} \rho_-, \text{ for } \bar{\rho}(t) < \rho^*\\ \rho_+, \text{ for } \bar{\rho}(t) \geq \rho^*.
\end{cases}
\end{eqnarray}
Here $\rho^*$ is a limiting density beyond which $\rho_l$ is reduced in order to control the average density $\bar{\rho}$. Note that all densities are normalized to remain in the interval $[0;1]$. For numerical simulations we chose an initial linear profile $\rho(x,0) = (\rho_r-\rho_-)x/L + \rho_-$ and let the system evolve into the steady state. We emphasize that the phase boundary between HD$_+$ and HD$_-$ phases depends on the initial condition.
\begin{figure}[h]
\centering
\subfigure[{} $\rho_- = \rho_+ = \rho_l$]
{\includegraphics[width=6.2cm]{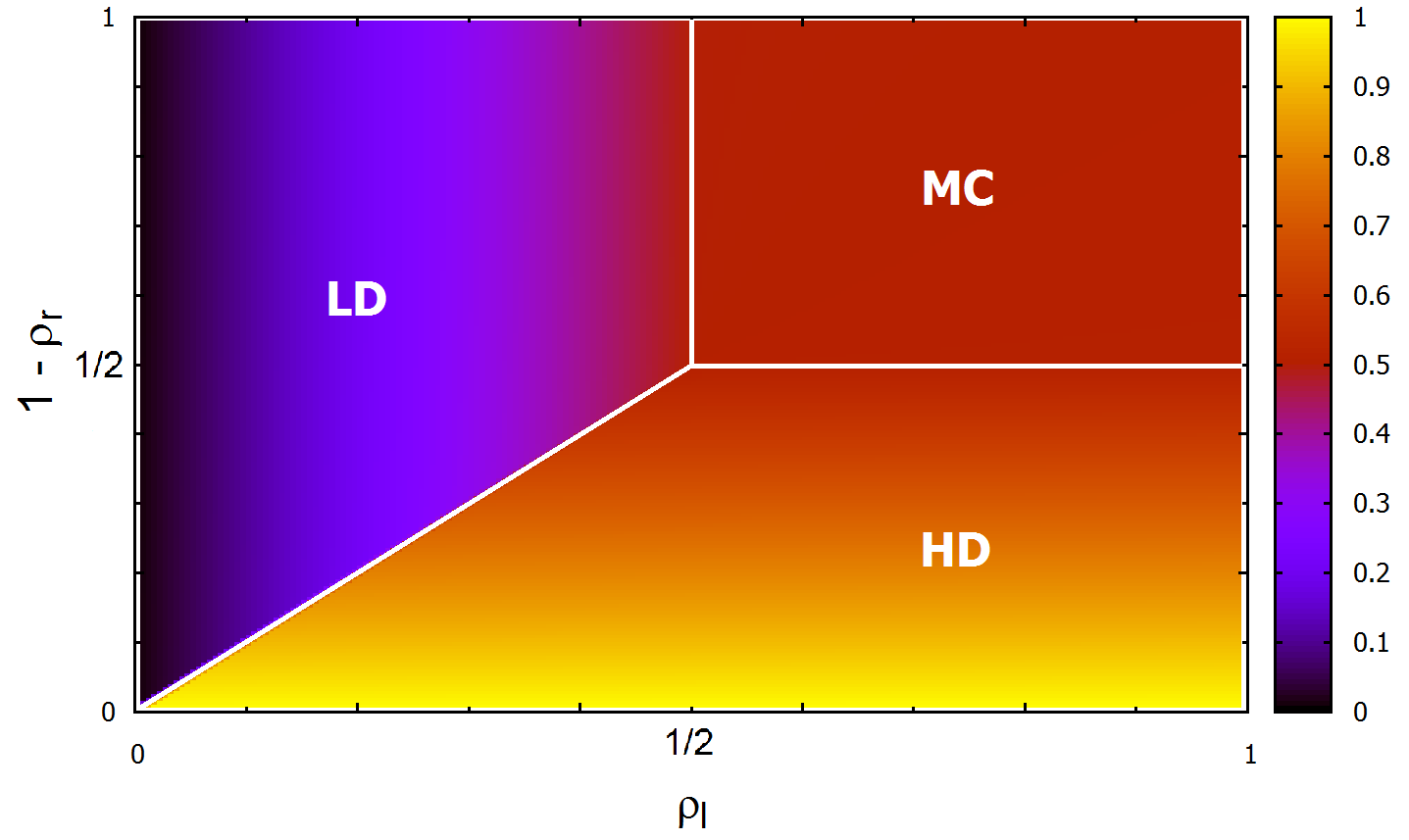}}
\subfigure[{} $\rho_+ < 1/2 < \rho_-$]
{\includegraphics[width=6.2cm]{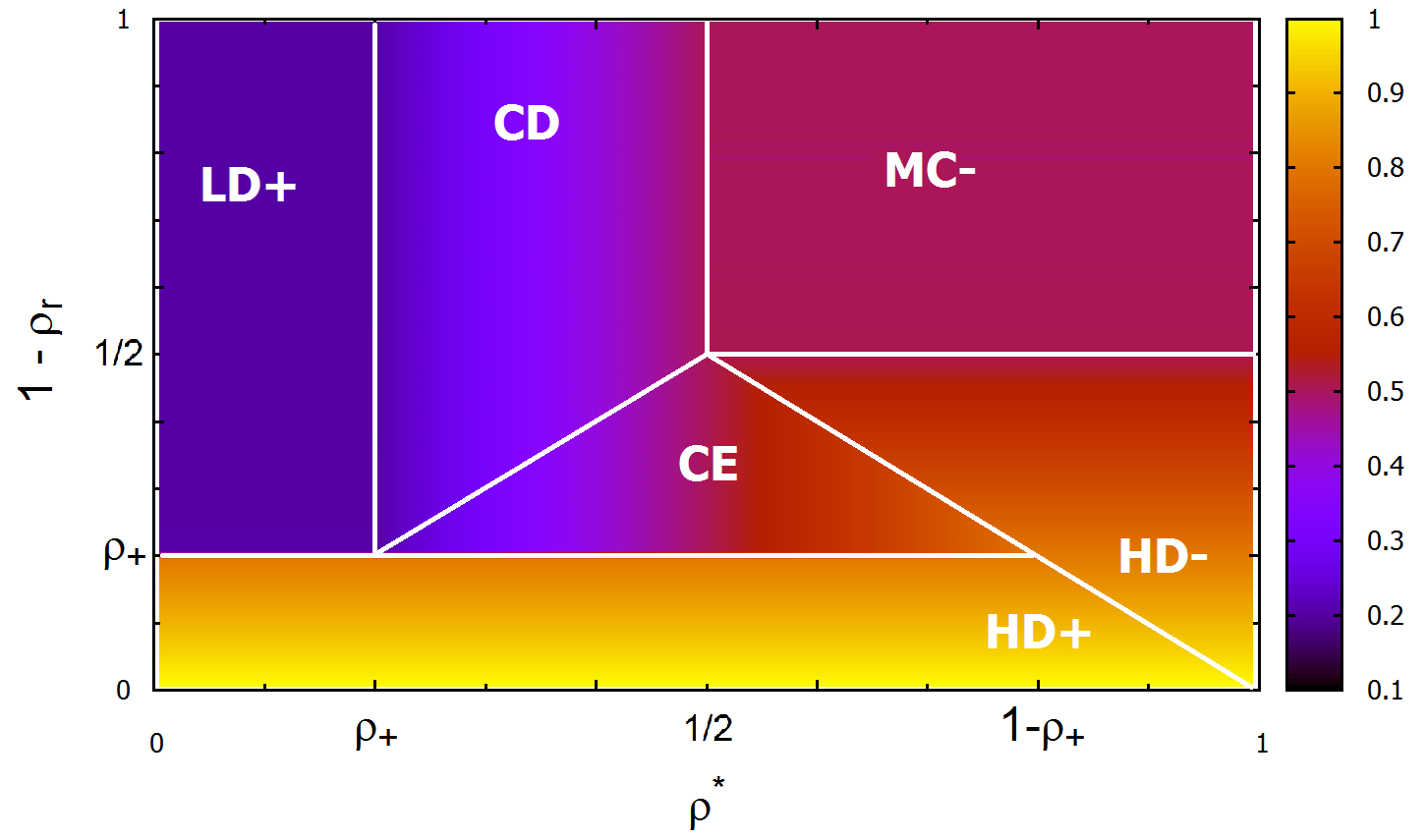}}
\subfigure[{} $1/2 < \rho_+ < \rho_-$]
{\includegraphics[width=6.2cm]{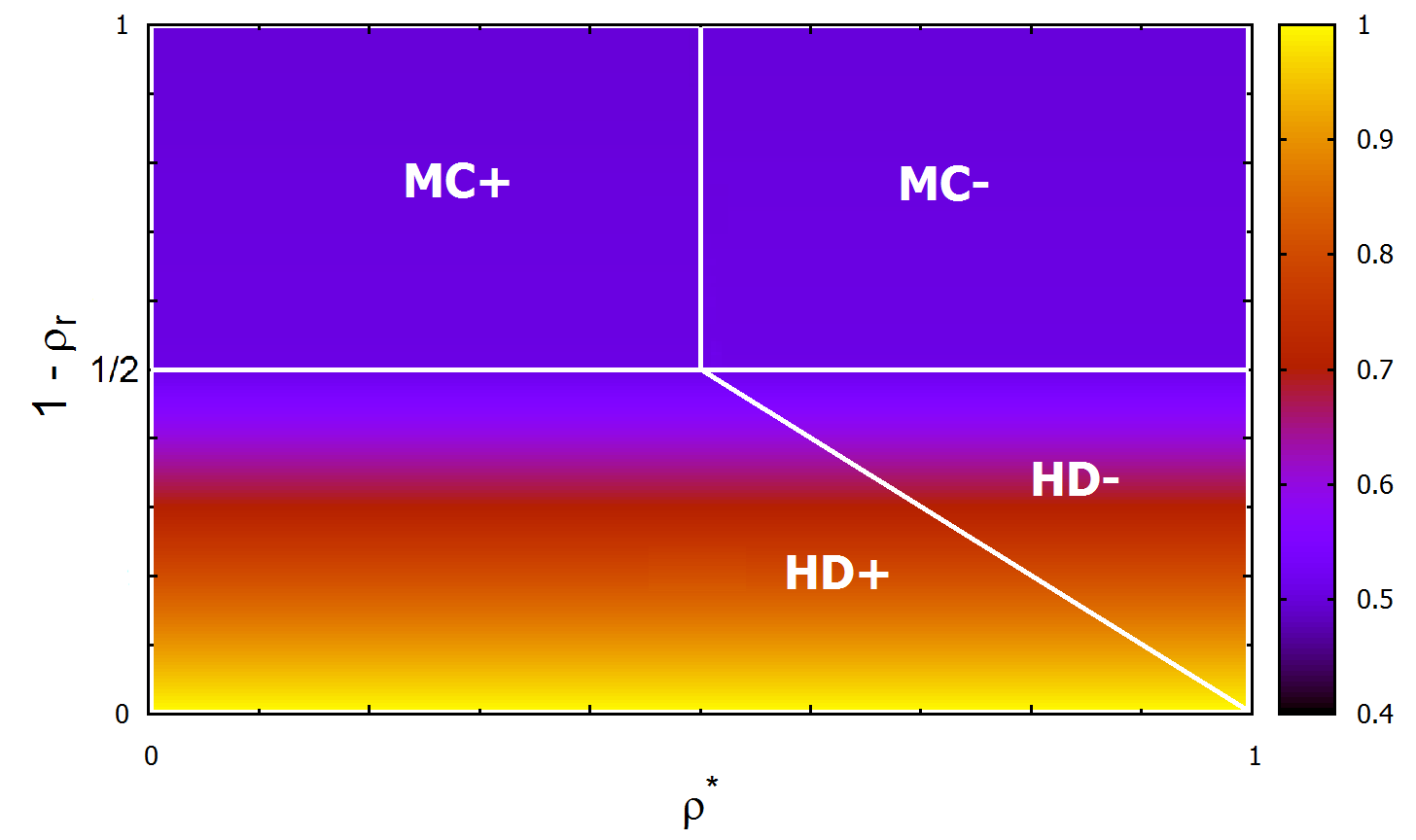}}
\subfigure[{} $ \rho_+ < \rho_- < 1/2$]
{\includegraphics[width=6.2cm]{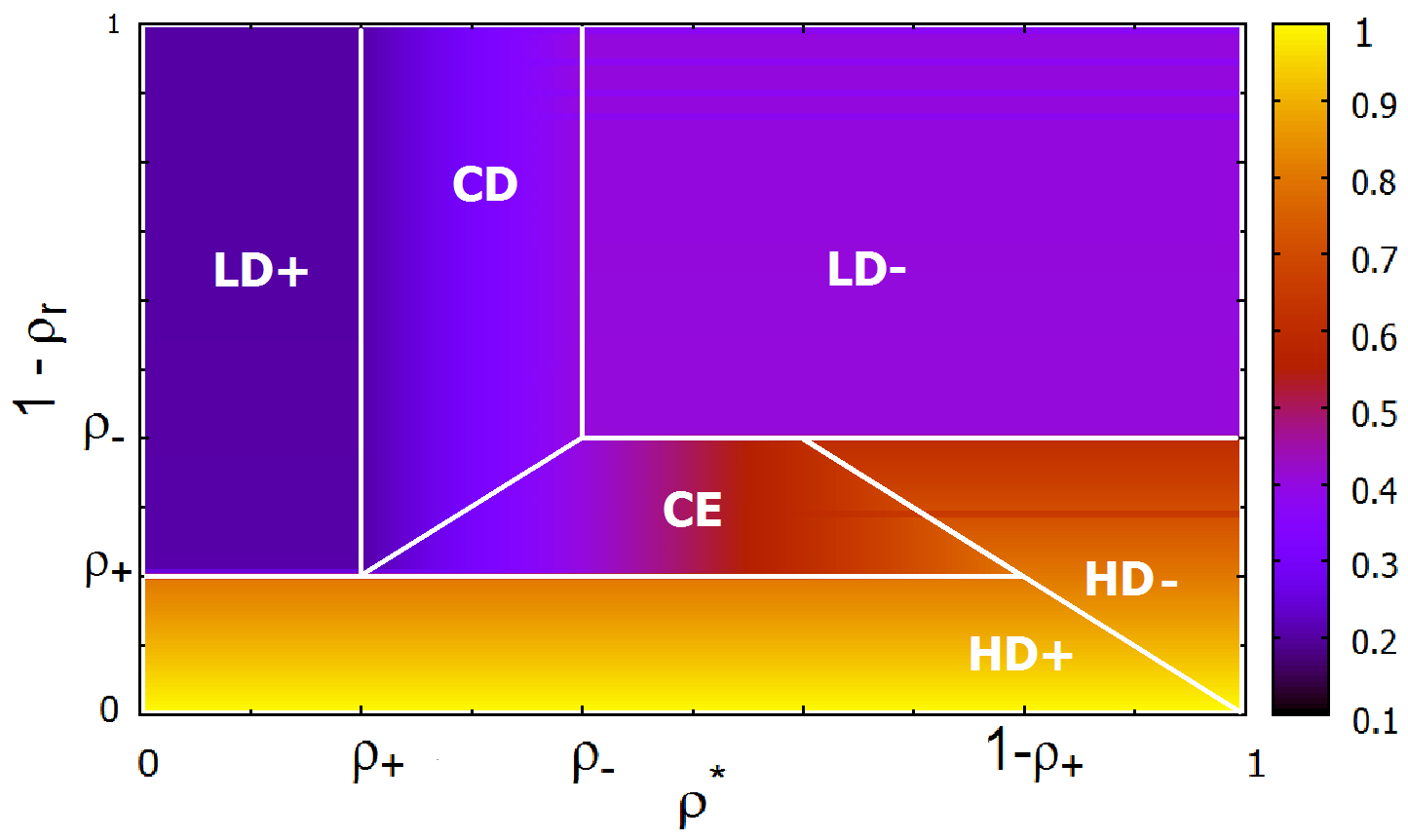}}
\caption{Phase diagram for (a) generic left b.c. (\ref{generic_bc}) and (b)-(d) dynamic b.c. (\ref{dynamic_bc}). The coloring encodes the value of the average density $\bar{\rho}$.
In (b): $\alpha_-=0.6$ and $\alpha_+=0.2$, in (c): $\alpha_-=0.8$ and $\alpha_+=0.6$, in (d) $\alpha_-=0.4$ and $\alpha_+=0.2$.}
\label{pha_dia}
\end{figure}
The numerical results for Burgers equation are obtained by spacial discretization. This leads to \cite{Pop}:
\begin{equation}
\frac{\partial}{\partial t}\rho_i = -(1-2\rho_i)\frac{\rho_{i+1}-\rho_{i-1}}{2} + D(\rho_{i+1}+\rho_{i-1}-2\rho_i).
\end{equation}
In the remainder of the paper the diffusion constant is set to $D=1/2$. Then this equation turns into
\begin{equation}
\label{MF}
\frac{\partial \rho_i}{\partial t} = \rho_{i-1}(1-\rho_i) - \rho_i(1-\rho_{i+1})
\end{equation}
which is nothing but the mean-field equation for the microscopic dynamics of the TASEP. The following section presents the results from numerical solutions of those mean-field equations.
\section{Analytic results}
\label{s2}
The mean-field theory assumes that correlations between neighboring sites vanish, so that the probability to find a certain lattice configuration factorizes into simple on-site factors, namely $\rho_i$ if site $i$ is occupied and $1-\rho_i$ if site $i$ is empty, compare \cite{Blythe}. In the present realization the boundary conditions are $\rho_{L+1} = \rho_r = \text{const}$ and
\begin{eqnarray}
\rho_{0} = \rho_l = \begin{cases} \rho_-, \text{ for } \bar{\rho} < \rho^*,\\
\rho_+, \text{ for } \bar{\rho} \geq \rho^*,\end{cases}
\text{with } \bar{\rho} = \frac{1}{L}\sum\limits_{i=1}^{L}\rho_i.
\end{eqnarray}
Further:
\begin{equation}
\rho_1(1-\rho_2)=\rho_l(1-\rho_1), \qquad \text{and } (1-\rho_r)\rho_L=\rho_{L-1}(1-\rho_L).
\end{equation}
The general solution for $1<i<L$ is \cite{Derrida92}
\begin{equation}
\label{rho_i}
\rho_i = \frac{-\rho_s\rho_u(\rho_s^{i-1}-\rho_u^{i-1})+(\rho_s^{i}-\rho_u^{i})\rho_1}{-\rho_s\rho_u(\rho_s^{i-2}-\rho_u^{i-2})+(\rho_s^{i-1}-\rho_u^{i-1})\rho_1}.
\end{equation}
Here $\rho_s$ and $\rho_u$ are the solutions of $J=\rho(1-\rho)$.
From FIG. \ref{pha_dia} (a) we can identify the well-known phases: {Low-density (LD) phase:} $\bar{\rho} = \rho_l$, for $1-\rho_r > \rho_l$ and $\rho_l < 1/2$, {High-density (HD) phase:} $\bar{\rho} = \rho_r$, for $1-\rho_r > \rho_l$ and $1-\rho_r < 1/2$, and {Maximum current (MC) phase:} $\bar{\rho} = \frac{1}{2}$, for $\rho_l, 1-\rho_r > 1/2$.
Now we investigate the new boundary condition (\ref{dynamic_bc}). TABLE \ref{tab1} shows the phases that can be identified.
\begin{table}[h]
\begin{tabular}{p{4cm}|p{1.7cm}|p{1.7cm}}
Phase                    & $\rho_l^{\text{eff}}$ & $\bar{\rho}$\\\hline
Low-density  (LD$_+$)    & $\rho_+$ &$\rho_+$\\\hline
Low-density  (LD$_-$)    & $\rho_-$ &$\rho_-$\\\hline
High-density (HD$_+$)    & $\rho_+$ & $\rho_r$ \\\hline
High-density (HD$_-$)    & $\rho_-$ & $\rho_r$ \\\hline
Maximum-current (MC$_-$)   & $\rho_-$ & $1/2$ \\\hline
Maximum-current (MC$_+$)   & $\rho_+$ & $1/2$ \\\hline
Controlled-density (CD)  & $\rho^*$ & $\rho^*$\\\hline
Co-existence (CE) phase  & $1-\rho_r$ & $\rho^*$ \\\hline
\end{tabular}
\caption{\label{tab1} Average left-hand- and overall density in the various phases.}
\end{table}
\begin{figure}[h]
\centering
\subfigure[{} CD-phase]
{\includegraphics[width=3.9cm]{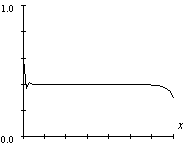}
\includegraphics[width=3.9cm]{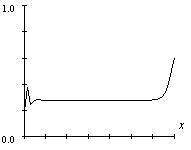}}
\subfigure[{} CE-phase]
{\includegraphics[width=3.9cm]{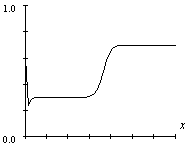}
\includegraphics[width=4.1cm]{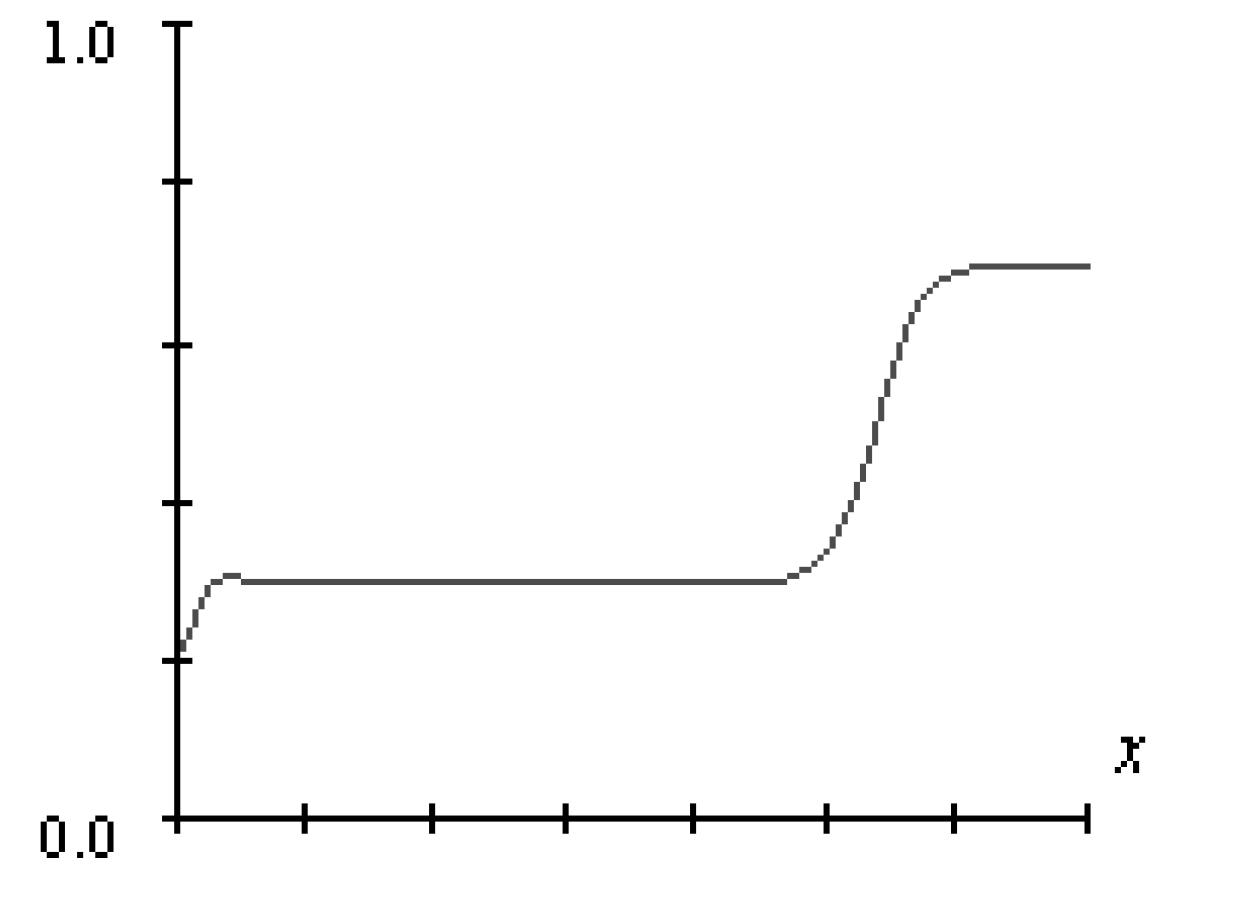}}
\caption{Density profiles for typical values of $\rho_l$ and $\rho_r$.}
\label{prof_mit}
\end{figure}
Before we turn into details, we emphasize that the various phases in TABLE \ref{tab1} indicated by $-$ and $+$ are coupled effectively by either of the left reservoirs at densities $\rho_-$ and $\rho_+$ respectively. Additionally, two phases are observed that are completely new compared to the generic TASEP, see TABLE \ref{tab1}. Those are the controlled-density (CD) phase and the co-existence (CE) phase. FIG. \ref{prof_mit} shows typical density profiles of those phases. One sees that the CE phase exhibits a stable upward shock that separates a high-density and a low-density region. In both phases the system is not dominated by contact with either of the two left reservoirs but both reservoirs are coupled in rapid alternation to the system. Summarizing, the stationary system behaves as if it would be coupled to an effective left boundary reservoir with constant density $\rho_l^{\text{eff}}$ that differs from phase to phase, see TABLE \ref{tab1}. In each phase, it is helpful to have in mind where on the horizontal axis of the generic phase diagram from FIG. \ref{pha_dia} (a) the values of $\rho_-$, $\rho_+$ and $\rho_l^{\text{eff}}$ locate. Then one can imagine in each case which phases are reached by variation of $\rho_r$, i.e.\ by moving vertically through the generic phase diagram. The reader shall imagine those vertical lines for $\rho_-$ and $\rho_+$ in order to understand phenomenologically the value of $\rho_l^{\text{eff}}$ in the different cases shown in FIG. \ref{pha_dia} (b) - (d) that are explained in the following.
We begin with FIG. \ref{pha_dia} (c): If both $\rho_+$ and $\rho_-$ exceed $1/2$ both those lines cross the MC-HD transition line. In both cases, MC and HD phases appear for $1 - \rho_r$ greater or smaller than $1/2$ respectively. In the MC phase, for $\rho^* < 1/2$ ($\rho^* > 1/2$), the average density $\bar{\rho} = 1/2$ is smaller (greater) than $\rho^*$. Therefore $\rho_l^{\text{eff}}$ equals $\rho_+$ ($\rho_-$) for $\rho^* < 1/2$ ($\rho^* > 1/2$) and the MC phase is distinguished in MC$_+$ and MC$_-$. Also the HD phase is distinguished further: Both (sub-)phases are separated by the line $\rho_r = \rho^*$. Since $\rho_r$ is the bulk density, in the region $\rho_r < \rho^*$ one finds $\rho_l^{\text{eff}} = \rho_-$ with the help of (\ref{dynamic_bc}). Therefore this is the HD$_-$ phase. Similar arguments hold for the HD$_+$ phase.\\
If $\rho_+ < 1/2$ and $\rho_- > 1/2$ one arrives at the phase diagram \ref{pha_dia} (b). The location of the LD$_+$ phase is explained as follows: First, from the TASEP phase diagram FIG. \ref{pha_dia} (a) it is known that a low-density state is reached for $1 - \rho_r > \rho_l$; second, if $\rho^* < \rho_+$ then it is evident from (\ref{dynamic_bc}) that the system behaves as if there would be a left boundary reservoir with density $\rho_l^{\text{eff}}=\rho_+$. In case of FIG. \ref{pha_dia} (b) only $\rho_-$ is large enough to lead to an MC phase. Hence the occuring phase has $\rho_l^{\text{eff}} = \rho_-$ and is referred to as MC$_-$ phase. The imaginary vertical line $\rho_+$ in FIG. \ref{pha_dia} (a) crosses the co-existence line between high- and low-density phases where $\rho_l=1-\rho_r$ in the generic TASEP. This crossing leads to the CE phase, consequently with $\rho_l^{\text{eff}} = 1 - \rho_r$. The CD phase is in fact a low density phase with $\rho_l^{\text{eff}}=\rho^*$, appearing here for $\rho_+ < \rho^* < 1/2$. What happens is quite intuitive: the system is equilibrated at the left end due to permanent change of contact with reservoir densities $\rho_+$ and $\rho_-$ around the control value $\rho^*$. In the same way one can explain the phase diagram FIG. \ref{pha_dia} (d). Just a remark on the appearance of the LD$_-$ phase: If $\rho^* > \rho_-$ it is expected with (\ref{dynamic_bc}) that the average density becomes $\rho_-$ and the system remains in contact with the $\rho_-$ reservoir. Finally we stress that the bulk density $\bar{\rho}$, given in TABLE \ref{tab1} can be deduced from the maximum-current principle \cite{Blythe},\cite{Pop} which takes here the form:
\begin{eqnarray}
      J = \bar{\rho}(1-\bar{\rho}) = \begin{cases} \text{min}_{[\rho_l^{\text{eff}},\rho_r]}\rho(1-\rho), \text{ if } \rho_l^{\text{eff}} < \rho_r,\\ \text{max}_{[\rho_l^{\text{eff}}, \rho_r]}\rho(1-\rho), \text{ if } \rho_l^{\text{eff}} > \rho_r,\end{cases}
\end{eqnarray}
In the CE-phase one finds co-existence of an HD phase at density $\rho_r$ and a CD phase at density $1-\rho_r$. Where both regions merge a shock is formed, see FIG. \ref{prof_mit}.
Since the average density remains $\bar{\rho}$ the position $x_s$ of the shock is given by $\rho^* = (1-\rho_r) x_s + \rho_r(L-x_s)$.
The phase diagram as depicted in FIG. \ref{pha_dia} (b) obviously holds only if we take $\rho_- > 1/2$ and $\rho_+ < 1/2$. If both values exceed $1/2$ the system is in HD-phases for $\rho_r>1/2$ and MC phases otherwise \footnote{Note that one special case that has completely different physics is excluded here: $\rho_-<\rho_+$.}. If both $\rho_+$ and $\rho_-$ have values below $1/2$ then obviously MC phases are suppressed. The results are shown in FIG. \ref{pha_dia} (c) - (d).
\section{Flow optimization by DFC}
\label{s3}
\subsection{Optimal choice of $\rho^*$}
The phase diagram of the TASEP with DFC (see FIG. \ref{pha_dia}) and the values of $\bar{\rho}$ in the various phases (see table \ref{tab1}) give an idea how to set the threshold $\rho^*$ in order to keep the flow as large as possible.  One can think of $\alpha_-$ being given by the (constant) demand of incoming drivers and $\beta$ being given by the characteristics of the outflow region of the bottleneck. We consider the scenario of FIG. \ref{pha_dia} (b) and thus argue from the viewpoint of the mean-field description. We move through the phase diagram on a virtual horizontal line for constant $\beta$. Here one can distinguish the following three cases: The bulk density starts at $\alpha_+$, then takes the value of $\rho^*$ and increases until it reaches the value of $1/2$ (case 1: for $1/2 < \beta < 1$) or $1-\beta$ (case 2: for $\alpha_+ < \beta < 1/2$). In case 3 (for $0 < \beta < \alpha_+$) the bulk density remains at $1-\beta$ for all choices of the threshold $\rho^*$. From the traffic viewpoint the interest is in maximising the flow. The closer the density is to $1/2$, the higher becomes the flow, due to the relation $J=\rho(1-\rho)$. Thus in case 1 the flow is maximized for $\rho^*\geq 1/2$ and in case 2 for exactly $1/2$ (in case 3, remember, is independent of $\rho^*$). Now consider FIG. \ref{pha_dia} (c). In case 1 the flow is maximised for $\rho^*\geq \rho_r$ ($=1-\beta$) and in case 2 for $\rho^*=1/2$. Finally consider FIG. \ref{pha_dia} (d). For $\beta>1/2$ ($\beta<1/2$) the flow is independent of $\rho^*$ equal to $1/2$ ($\beta(1-\beta)$). Thus, concluding one can say that the choice $\rho^*=1/2$ theoretically is always the best in order to maximise the flow. This result is expected since this is the density at which the flow has its maximum. Therefore in the following we restrict ourselves to this case, noting that results easily convert to the general case.
\subsection{Benefit by DFC}
\begin{figure}[h]
\centering
\subfigure[{} Optimization of density and flow by DFC]
{\includegraphics[width=9.2cm]{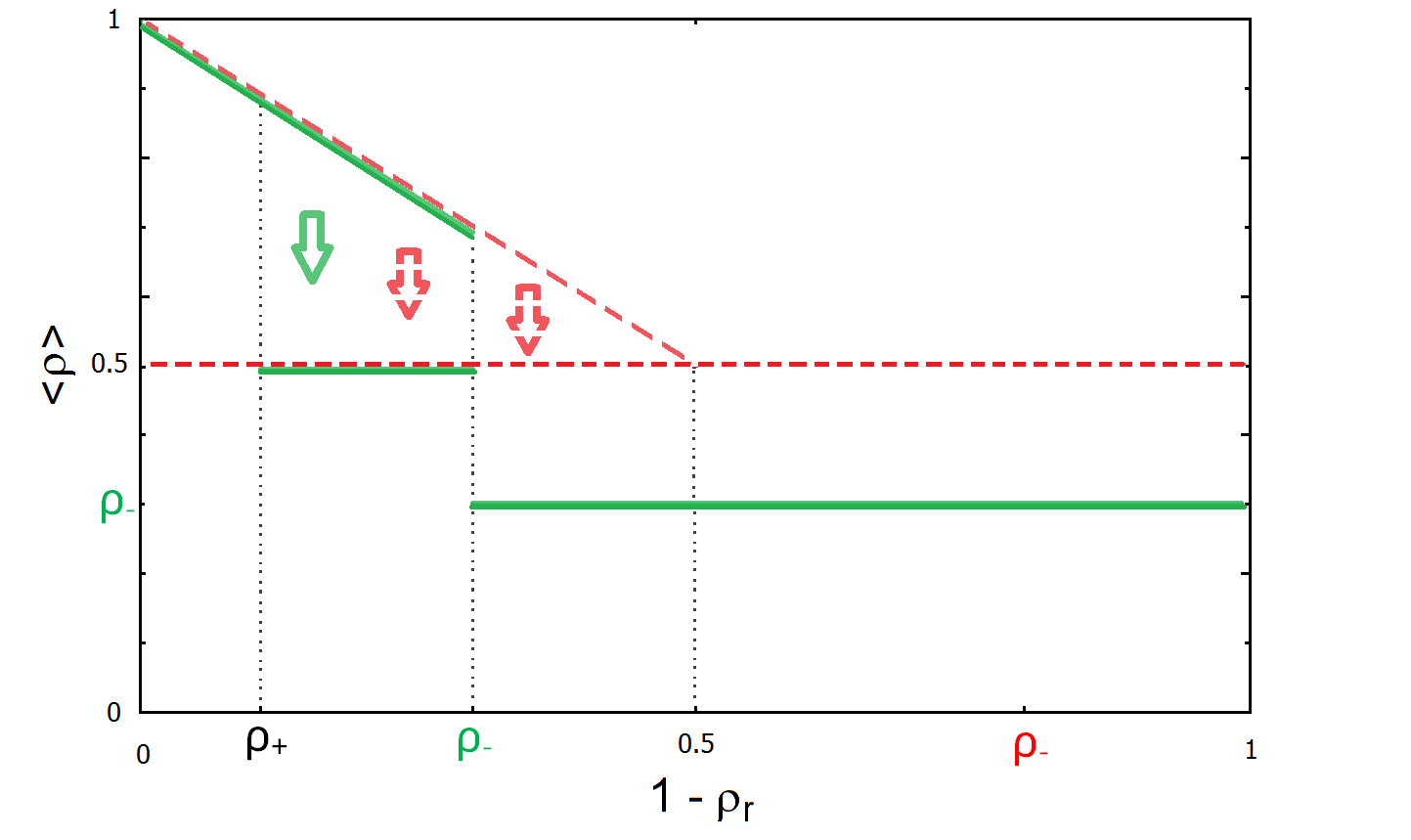}}
\subfigure[{} Phase diagram showing benefit of DFC]
{\includegraphics[width=9.2cm]{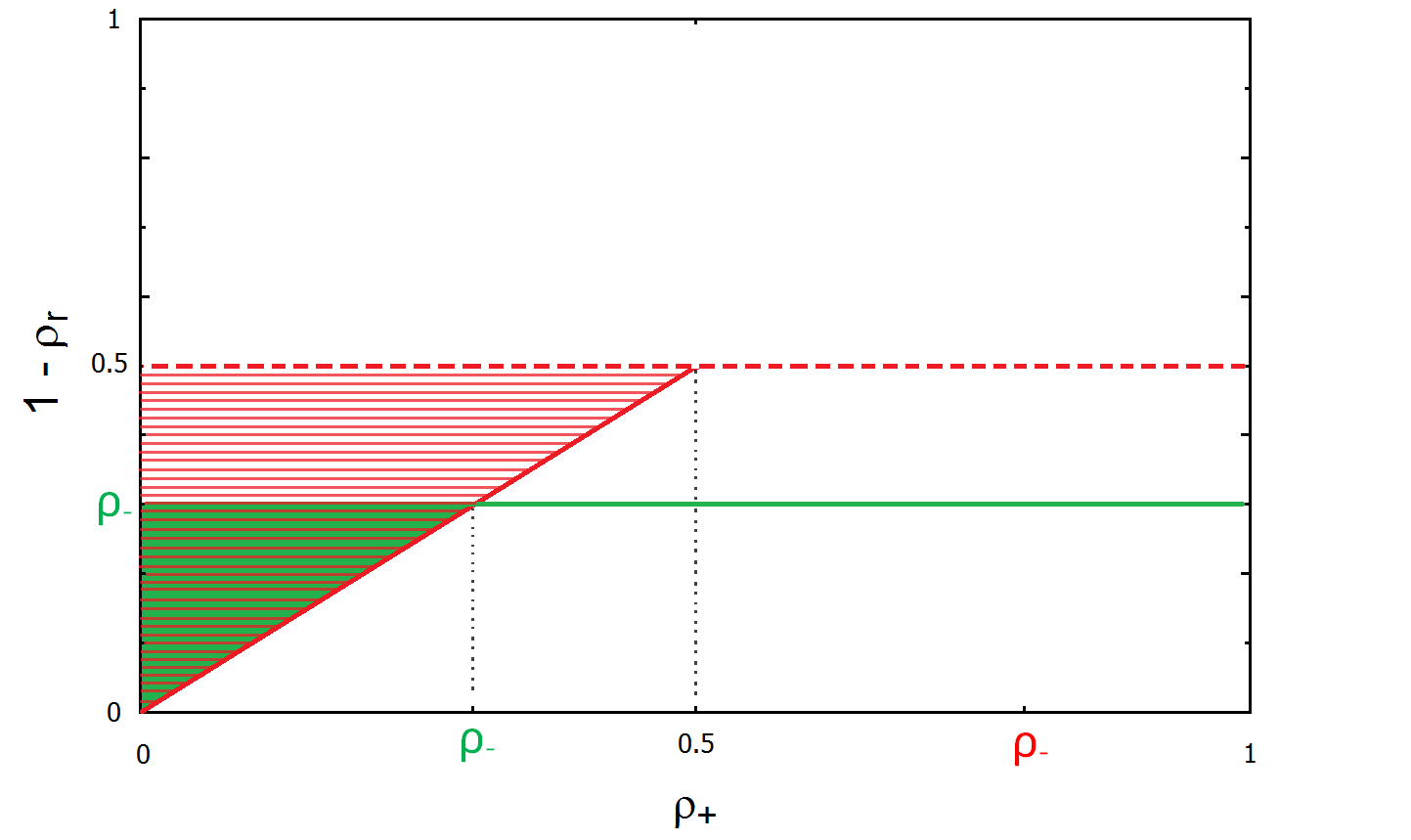}}
\caption{(a) Conversion of high density into density 1/2 by DFC. (b) The dashed (continuous) triangle is the region that is optimized to a maximum-current region by density-feedback control for $\rho_- > 1/2$ ($\rho_-<1/2$).}
\label{fazit}
\end{figure}

FIG. \ref{fazit} illustrates the benefit of DFC. Dashed (continuous) objects correspond to the case where $\rho_->1/2$ ($\rho_-<1/2$). In FIG. \ref{fazit} (a) we draw an analogy to the generic system in assuming that $\rho_-$ corresponds to the generic left reservoir density. Then the figure shows that the switching to a lower density $\rho_+$ leads to a conversion of a high density to density $1/2$. FIG. \ref{fazit} (b) shows the benefit of DFC in the $(\rho_+,\beta)$-plane. For simplicity we write $\beta$ instead of $1-\rho_r$. One sees the according additional triangular MC region belonging to this benefit.
Above the dashed line (and $\rho_->1/2$) the system is in MC phase. The outflow is high enough ($\beta>1/2$) to suppress HD phases and therefore no optimization is possible there. Similarly, above the continuous line (and $\rho_-<1/2$) the system is in LD phase where the density is smaller than $1/2$. Since DFC can only lower the density, here, the flow can never be optimized. To the right of the triangles ($\rho_+>\beta$) and below the whilst line ($\beta<1/2$ or $\beta < \rho_-$ respectively) one finds the HD phase. Since both $\rho_-$ and $\rho_+$ are larger than $\beta$, the inflow is always higher than the outflow and the high-density phase can not be left by variation of $\rho_+$.
\section{Simulation results}
\label{s4}
We repeat that the results of section \ref{s2} are exact consequences of the discretized Burgers equation (\ref{MF}), however they will, in general, not be exact for the corresponding TASEP with DFC, since the latter is described by (\ref{MF}) on a mean-field level. The weakness of the mean-field approach is that it ignores correlations arising from spatial inhomogeneities, including the existence of boundaries. However for the quantity of interest, namely the average density at threshold $\rho^*=0.5$ results will turn out to be in good agreement.
\subsection{Simulation of the TASEP with DFC}
\begin{figure}[h!]
\centering
\subfigure[{} HD$_+$ phase]
{\includegraphics[width=3.8cm]{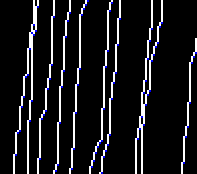}}
\subfigure[{} CE phase]
{\qquad\includegraphics[width=3.8cm]{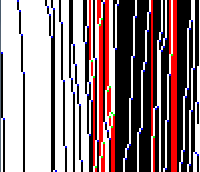}}
\subfigure[{} MC$_-$ phase (transition line to CD phase)]
{\includegraphics[width=3.8cm]{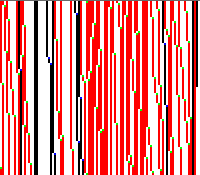}}
\subfigure[{} LD$_-$ phase]
{\qquad\includegraphics[width=3.8cm]{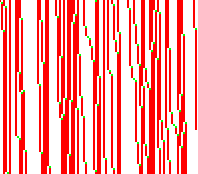}}
\caption{Space-time plots for $\rho^*=0.5$ in a system with $100$ cells. In (a)-(c): $\alpha_-=0.6$, $\alpha_+=0.2$ with  and $\beta=0.1$ (a), $\beta=0.3$ (b), $\beta=0.6$ (c). Further: $\alpha_-=0.4$, $\alpha_+=0.2$, $\beta=0.6$ in (d).}
\label{space_time}
\end{figure}

FIG. \ref{space_time} shows space-time plots with increasing space-coordinate in the right direction and time increasing in downwards direction.
Particles that entered at densities lower than $\rho^*$ are painted green (moving) and red (standing) while particles that entered at higher densities are colored blue (moving) and black (standing). Plotted are only those time steps where a move occurs. The subfigure captions give the name of the corresponding phase from mean-field.\\
In order to average quantities in the steady state, it turned out that the simulation of the TASEP with DFC converges very slowly. Therefore, as in \cite{Adams}, \cite{Cook08}, \cite{Cook}, it was chosen to feed the simulation at the expected density. For our studies thus the mean-field density serves as initial value. During $2\times 10^6$ time steps the system is let alone and afterwards every $100$ time steps the density is measured over $5\times 10^6$ steps. The average over the steady states of $100$ different initial configurations is taken.
\subsection{Comparison with mean field}
\begin{figure}[h!]
\centering
\subfigure[{} Density profiles]
{\includegraphics[width=8.7cm]{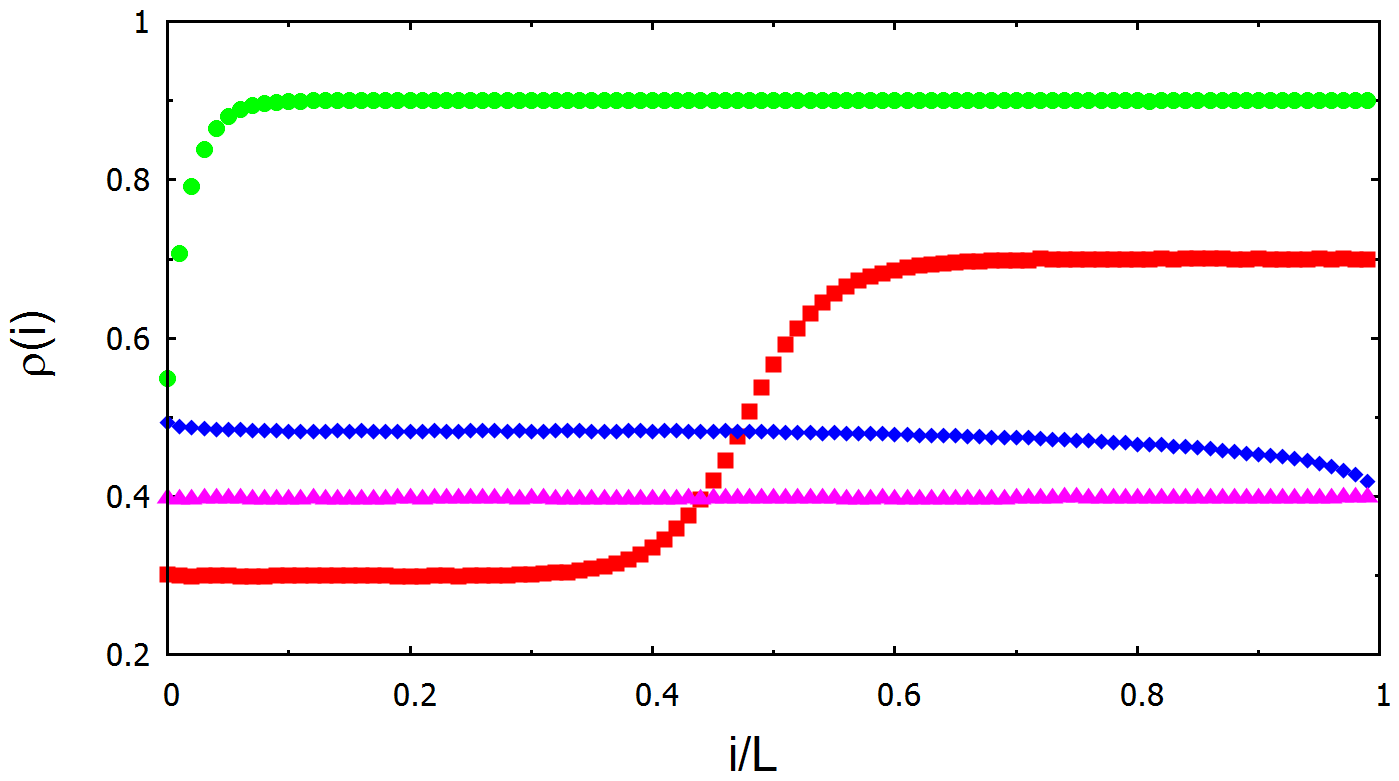}}
\subfigure[{} Average density]
{\includegraphics[width=8.7cm]{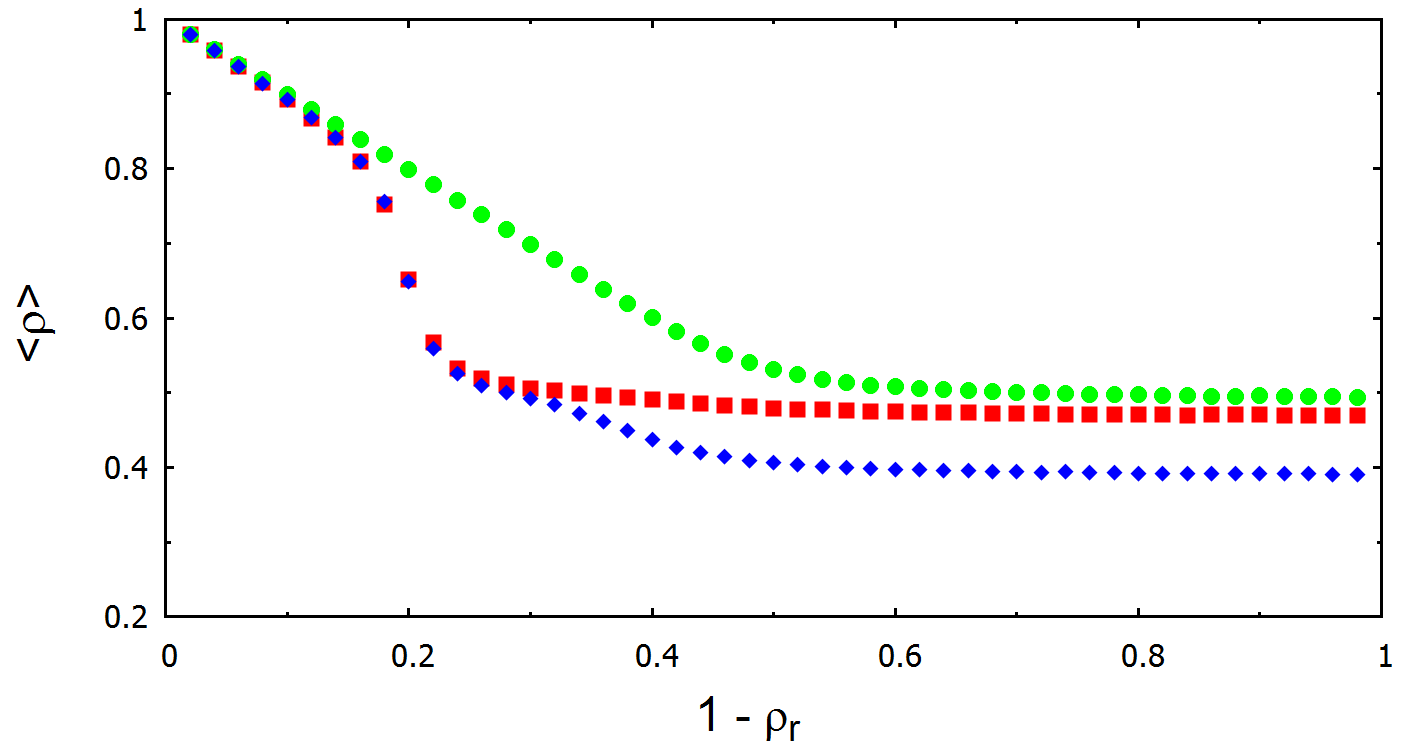}}
\caption{The figures show simulation results in case of $\rho^*=0.5$ for a system of length $100$: (a) Density profiles  corresponding to FIG. \ref{space_time}. (b) Average density versus $\beta$. Squares belong to $\rho_+=0.2$ and $\rho_-=0.6$ ( parameter case as in FIG. 1 (b)), circles correspond to $\rho_+=0.6$ and $\rho_-=0.8$ (parameter case as FIG. 1 (c)),  and diamonds correspond to $\rho_+=0.2$ and $\rho_-=0.4$ (parameter case as FIG. 1 (d)).}
\label{Dichte_beta}
\end{figure}

First, we will verify that the different phases resulting from the mean-field theory indeed occur in the TASEP with DFC and that the physics is correctly predicted. FIG. \ref{Dichte_beta} (a) shows the simulated density profiles that correspond to the space-time plots of FIG. \ref{space_time}: The circles saturating at density $0.9$ show the HD profile of FIG. \ref{space_time} (a) and reproduce the mean-field density $\rho_r$ of HD phases. The profile of squares corresponds to the CE phase of FIG. \ref{space_time} (b) and clearly shows the co-existence of low- and high densities so that the existence of the shock phase in the TASEP with DFC is verified. The profile corresponding to FIG. \ref{space_time} (c) on the transition line between CD and MC is given by the blue diamonds showing the flat profile around density $1/2$ which is the average density predicted by mean field. Finally the situation of FIG. \ref{space_time} (d) has the flat profile with constant density $0.4$. Note that in this case ($\rho_l^{\text{eff}}=1-\rho_r$) mean-field becomes exact.

Now we turn to the simulation of the average density in the system against $\beta$ in order to verify that densities and phase boundaries are correctly predicted. The results are shown in FIG. \ref{Dichte_beta} (b). See figure caption for more details. One sees that the circles are on the line $\rho = \rho_r$ for $\beta < 0.5$ and that $\rho = 0.5$ for $\beta \geq 0.5$ (which corresponds to the transition from HD to MC) as predicted by mean field. The squares start in HD and clearly jump at $\rho_+=0.2$ to density $1/2$ (corresponding to CE and MC). The diamonds clearly show three phases (as can be seen from FIG 1 (d)). Starting at HD one sees the kink at $\beta = \rho_+$ to the CE phase and another transition at $\beta=\rho_-$ to MC and $\rho=1/2.$ which is also in agreement with our mean-field presictions. Of course the sharpness of the transitions could be ameliorated by taking larger system sizes.

\section{Conclusion}
This paper studied a bottleneck situation of traffic with inflow at the left and outflow at the right end which was modeled by TASEP and Burgers equation. For this situation a concept to control the overall density has been analyzed. The left reservoir density takes the form $\rho_l(\bar{\rho}(t))$ and thus depends on the density at time $t$, generalizing the generic constant left reservoir density. It is reduced from $\rho_-$ to $\rho_+$ if the spatially averaged density $\bar{\rho}(t)$ at time $t$ lies above a certain threshold $\rho^*$. In contrast, the right end is kept in contact to a reservoir at fixed density $\rho_r$. The mechanism is referred to as density-feedback control (DFC). The same mechanism is provided in every-day life, where cars enter a dense road section at a smaller rate when there are possible alternatives. The paper showed that DFC can be efficiently used to maximize the flow by converting a fraction of the high-density phase to a maximum-current phase.\\
From numerical solution of the discretized Burgers equation the phase diagram in the plane spanned by $\rho^*$ and $1-\rho_r$ was derived that showed a rich phase behavior. The process exhibits two low-density, high-density and maximum-current phases that correspond to the two left boundary reservoirs. In addition there is a phase in which high and low-density co-exist so that a macroscopic shock profile can be observed. This phase corresponds to the co-existence line in the generic model between low- and high density phase. There also is a phase that is completely new compared to the generic model but can be anticipated intuitively; in this phase, the repeated change of the left-hand reservoir density around the threshold $\rho^*$ leads to an effective density $\rho^*$. It was further investigated for which choice of $\rho^*$ the flow is maximized. It could be shown that, although in the generic TASEP the flow is monotonically increasing with the left reservoir density, DFC optimizes the flow if the threshold density is chosen appropriately.\\
For the optimal choice of the threshold ($\rho^*=1/2$) we verified with the help of Monte-Carlo simulations that mean field correctly predicts the average density (and therewith the flow) in the system as well as the physics of the various phases including the co-existence phase. Note that simulations in which the Heaviside dependence of the density was replaced by a hyperbolic tangent with appropriate sharpness, inspired by \cite{Adams} have also been performed. This takes into account a (realistic) delay of the adjustment of the left density through feedback control. Further the model with parallel dynamics has been considered \cite{to_be_pub}. It turned out that results agree very much with the continuous-time case studied here. Further investigations could focus on the Nagel-Schreckenberg model of traffic flow. It is known that the phase diagram of the Nagel-Schreckenberg model remains even for larger maximum velocity \cite{Barlovic} (where cars can move more than a single site per timestep). While in the present model flow optimization is achieved at a threshold density $1/2$ one should decide whether this generalizes to the density at which the flow becomes maximal (as one would expect \cite{Markos2}). The next step is a generalization to more realistic microscopic traffic models as for example the Krau{\ss}-model \cite{krauss} in order to study effective traffic-management strategies based on DFC.

\end{document}